\documentclass{article}
\usepackage{amsmath,amsfonts,graphics,setspace}
\usepackage{chicago}

\newcommand{\egc}{\mbox{e.\,g.\,}}

\newcommand{\dr}[1]{\ensuremath{\mathrm{d} #1\,}}
\newcommand{\mc}[1]{\ensuremath{\mathcal{#1}}}

\newcommand{\ket}[1]{\ensuremath{\left|  #1 \right\rangle}}
\newcommand{\bra}[1]{\ensuremath{\left\langle #1 \right|}}

\newcommand{\proj}[2]{\ensuremath{\ket{#1} \bra{#2}}}

\newcommand{\matel}[3]{\ensuremath{\bra{#1} #2 \ket{#3}}}

\newcommand{\op}[1]{\ensuremath{\widehat{\textsf{\ensuremath{#1}}}}}
\newcommand{\opad}[1]{\ensuremath{\op{#1}^{\dagger}}}

\newcommand{\denop}{\ensuremath{\rho}}

\newcommand{\be}{\begin{equation}}
\newcommand{\ee}{\end{equation}}

\renewcommand{\denop}{\op{\rho}}

\begin{document}

\title{Quantum Mechanics on Spacetime I:\\ Spacetime State Realism}
\author{David Wallace \and Christopher G. Timpson}

\date{May 2009}

\maketitle


\begin{abstract}\noindent What ontology does realism about the quantum state suggest? The main extant view in contemporary philosophy of physics is \textit{wave-function realism}. We elaborate the sense in which wave-function realism does provide an ontological picture; and defend it from certain objections that have been raised against it. However, there are good reasons to be dissatisfied with wave-function realism, as we go on to elaborate. This motivates the development of an opposing picture: what we call \textit{spacetime state realism}; a view which takes the states associated to spacetime regions as fundamental. This approach enjoys a number of beneficial features, although, unlike wave-function realism, it involves non-separability at the level of fundamental ontology. We investigate the pros and cons of this non-separability, arguing that it is a quite acceptable feature; even one which proves fruitful in the context of relativistic covariance. A companion paper discusses the prospects for combining a spacetime-based ontology with separability, along lines suggested by Deutsch and Hayden
\end{abstract}

\section{Introduction}

In this paper we address two main questions: Suppose one were a realist about quantum mechanics; and suppose, in addition, that in the version of quantum mechanics one was committed to, the state is to be understood as standing for some thing in the world. How is one to think of what the world is like in that case? What \textit{is} it that the state stands for? This is our first question.

Our second follows on. Quantum mechanics is well-known for its nonlocal features; for its failure to satisfy a principle of separability. Quantum states might, generically, be non-separable; but given an answer to our first question, will it be true that the \textit{thing} thereby represented (however we are to think of it) is non-separable too? That is, when we are more careful to specify what the ontology associated with realism about the quantum state is, will that ontology satisfy separability, or not?

There has been remarkably little discussion of what the correct way of thinking about the ontology of the quantum state is. Here we develop an approach which we term \textit{spacetime state realism}, in opposition to \textit{wave-function realism}, which has so far been the default option; apparently for want of alternatives.

We begin (Section~\ref{prolegomena}) by sketching some requirements on specifying an ontology for quantum mechanics, and go on to introduce wave-function realism (Section~\ref{wf realism}), indicating how it meets these requirements. But there are good reasons to resist wave-function realism: it is plausible only in the restricted setting of fixed-particle number non-relativistic quantum mechanics and it has no reasonable extension to more physically realistic situations (Section~\ref{problems}). We therefore develop a position which avoids these problems, taking states of spacetime regions as the primary elements (Section~\ref{st ontology}); and we enumerate its benefits over wave-function realism.

What of separability (Section~\ref{locality})? The spacetime state ontology is unambiguously non-separable: the state of a given spacetime region will not be determined by the states of the sub-regions of which it is composed. By contrast, however, the wave-function ontology \textit{is} separable at the fundamental level. This marks a significant difference between these ontological pictures, but we go on to argue that the failure of separability is by no means a pathological feature of our preferred account. Rather, it is a feature which has active benefits when one comes to consider relativistic covariance and an important phenomenon recently noted by David Albert, that quantum mechanics exhibits so-called \textit{narratability failure} (Section~\ref{relativity}).

Our primary concern here is with realist accounts of quantum mechanics that involve purely unitary evolution, thus the Everett interpretation will loom large; however a number of our remarks will apply equally to other kinds of theories which take the quantum state seriously as an element of reality, such as Bohm-like hidden variable theories and dynamical collapse theories. We consider the matter of non-unitary dynamics within our spacetime state ontology in Section~\ref{nonunitary}.

In a companion paper \cite{part2} we explore the matter of separability in more detail; and investigate the remarkable possibility held out by \citeN{deutschhayden} that one may gain the benefits of a spacetime-based approach to quantum ontology while retaining separability at the fundamental level.

\section{Prolegomena}\label{prolegomena}

What ontology does quantum theory prescribe? Historically, finding an answer has not proven easy, for a variety of reasons. One way at least to set foot on a path to an answer is by beginning with the question of what in \textit{general} might be involved in specifying an ontology associated with a physical theory. We have no wish to be prescriptive, but it seems reasonable to agree that at least \textit{one} way in which an ontology may be specified (at least minimally) is by identifying suitable bearers for the properties that the theory trades in, and then going on to say something more about those properties themselves.
The austere mathematically framed description of these properties provided by theoretical physics might, on its own, seem a little thin: one would also like to be in a position to say something about how they relate to the everyday world of moderately sized objects which surrounds us.
It is, however, notoriously difficult to provide a satisfactory story about any of these various elements when discussing quantum theory. Potential bearers and properties that one might identify generally seem too rebarbatively alien to get a good grip on; and the problem of measurement casts a long shadow over the question of recovering appearances (see \citeN{wallaceQMreview} for a review).

But what might be required in order to gain at least a reasonable grip on a proposed ontology? We should be be wary of setting our sights too high and insisting on too much when trying to ponder what fundamental physics presents us with. Consider, for example, a simpler case: that of electromagnetism.

We are happy enough that electromagnetism does not present us with an irredeemably obscure picture of the world, yet it is not as if we really have an intuitive grasp
of what an electric or a magnetic field is, other than indirectly and by means of instrumental
considerations (``A test charge would be accelerated \textit{thus}.", for example). But in the case of field theory, we do at least understand that ``the electric field
at spacetime point $p$'' denotes a property \emph{of point p},\footnote{Strictly speaking, assigning a vectorial quantity to a point will also involve making reference to relations with neighbouring points, c.f. \citeN{butterfield:pointillisme}.} and this gives us substantial intuitive
understanding of how to think about that field: namely, it describes certain, admittedly somewhat alien,
properties \emph{of spacetime regions}.

Does electromagnetism offer any \textit{further} elucidation of its ontology, beyond an assignment of vectorial quantities to spacetime points? We think not: what could that understanding derive from? All we have to work with are the mathematical structure of the theory and the instrumental considerations alluded to above---and the latter, when one takes seriously the idea of physics as universal, ultimately collapse into interactions between the electromagnetic field and other comparably alien entities. (It may help here to recall that \emph{matter}, in the sense of the macroscopic material bodies we observe, consist of, and manipulate in the lab, cannot ultimately be thought of as some non-electromagnetic entity on which electromagnetic forces act: modern physics makes clear that solid matter is made up of electromagnetic fields as well as fermionic matter. In any case, the latter is scarcely less alien than the former. Even if we naively regard fermions as classical point particles, ``point particles'' are shapeless, colourless, textureless entities with little in common with familiar macroscopic bodies, for all that schoolbook physics may have acclimatised us to them.)

Thus it seems that we gain a basic understanding of the electromagnetic field by seeing it as a property of spatial regions\footnote{We do not wish, in making this claim, to make some sort of dogmatic commitment to spacetime substantivalism. It may well be, for instance, that talk of spatial \emph{regions} is in principle eliminable in turn in favour of talk of the spatial \emph{relations} between events, as in Barbour's version of field theory \cite{barbourrelativitywithoutrelativity}.}; and our further understanding must be mediated by reflecting on its role in the theory (including, importantly, the instrumental considerations); beyond that there doesn't seem to be much further to be grasped.

Now consider a converse example. In \emph{classical} mechanics (of particles or fields) it is possible to construct very elegant and abstract formulations in terms of
trajectories on a symplectic manifold. Such formulations are normally understood as a mere mathematical convenience---a shorthand for the real field-based or particle-based description---but how should we think about such a theory (the theory, say, of $10^{23}$ point particles) if we were given it \textit{without being told} that it was intended in that way?

The theory would certainly look alien. It would present a single point evolving in an enormously high-dimensional ($6 \times 10^{23}$-dimensional) symplectic space. The space should not be understood as featureless, though (or even featureless other than the \emph{symplectic} structure): the Hamiltonian that guides the point's evolution will be highly non-invariant under generic symplectomorphisms (transformations only preserving symplectic structure), and so the symmetry group of the theory will be very small compared to the symplectic group. On the usual assumption (see, e.\,g.\,, \citeN{earman}) that a spacetime must have a structure rich enough to formulate its dynamics, but no more\footnote{The conclusion also follows from Brown's thesis \cite{harveyref} that a spacetime's structure is just a \emph{description} of its dynamics.}---and so can have no invariants or non-invariants under transformations that the dynamics do not have---this means that the space (unlike, say, $6 \times 10^{23}$-dimensional Euclidean space) is highly non-isotropic and non-homogeneous.

Nonetheless, if we were to think of our theory as just about the evolution of a single particle on that huge, exotically structured space, we would have failed to understand its structure fully. True understanding comes from the recognition that the theory is naturally isomorphic to---and so can be understood as representing---the positions and velocities of $10^{23}$ particles in ordinary, Euclidean, three-dimensional space. Only once we recognise that fact can we get a clear grip on the theory's structure, or relate it to our observed world of spatially located objects.

Our claim, in essence, is that thinking about quantum mechanics in terms of a wavefunction on
configuration space is rather like thinking about classical mechanics in terms of a point on phase space. In both cases, there is a far more perspicuous way to understand the theory, one which is connected to spacetime in a more direct way.

Before moving on, a caveat is in order.
In trying to address the question of how the world is according to quantum mechanics, we wish, so far as possible, to remain neutral between a metaphysically substantive and a deflationary reading; neutral, that is, between the (substantive) thought that we can successfully talk about the one correct reading of the quantum state (description of how the world is if QM is true) and the (deflationary) thought that perhaps the best we can do is compare the merits of different more or less perspicuous descriptions of something we can get no closer to (while remaining within the ambit of the theory). Thus while we argue for the adoption of spacetime state realism over wave-function realism, we wish to remain neutral on whether one of these (or perhaps some third) really does provide the One True Interpretation of the quantum state, or whether one is merely a more perspicuous description than the other; a description of something that we are ultimately unable to render unequivocally in intuitive terms.

Equally in classical physics, it seems clear to us that the more perspicuous description of the theory is in terms of $N$ particles in three-dimensional space, not one particle in $6N$-dimensional space, but we are neutral (here, at any rate!) as to whether this means that the $N$-particle story is closer to the truth, or just that the $N$-particle story is a more sensible way of describing the structure of the world.

We do, however, want to stress that we regard the one-particle story as a \emph{coherent possibility} (whether or not it is just the same possibility as the $N$-particle story, differently described)---and that if the $N$-particle story were empirically adequate (which it isn't, of course) then so would the one-particle story be. For that story (by construction) is isomorphic to the $N$-particle story, and the emergence or otherwise of higher-level ontology from lower-level theories depends, to our minds, primarily on the structure of those theories and not on their underlying true nature (whatever that is). On the one-particle theory, three-dimensional space would turn out to be emergent, but it would be no less real for that: indeed, it would be emergent in such a strong and exact sense as to make us inclined to regard the $N$-particle story as to be preferred.\footnote{We have found, in conversation, that there are some for whom it is just intuitively obvious that the world isn't like that. But personally speaking, we have no pre-theoretic intuitions at all about what it would be like if the world were a highly non-isotropic $6 \times 10^{23}$-dimensional space in which one particle moved, and we wouldn't trust such intuitions if we did.}

One might put things in these terms: in our view, there is no guide to the ontology of a mathematically formulated theory beyond the mathematical structure of that theory (including that structure to do with observable quantities and experiment). But when trying to learn ontological lessons from the theory, one does well to prefer a representation which makes manifest the structure that the theory ascribes to the world. One of the difficulties with the single particle, high dimension story about classical mechanics is that it makes it less immediately obvious what structure is being attributed to the world. The physical world is very richly structured; a distribution of field amplitudes over 3-space, or an arrangement of particles in 3-space is very obviously richly structured; a single point in high dimensional space less obviously so. Despite the structure to that space implied by the Hamiltonian, one might conceivably miss the isomorphism to $N$ 3-space particles.

\section{Wave-function realism}\label{wf realism}

Let us now consider how we might go about understanding the quantum state.
A traditional view is that the state should be understood as somehow being a description or encoding of various classically describable or measurable properties. According to this view, the potential properties of the system are represented by projectors on Hilbert space (although typically not all such projectors will correspond conversely to \textit{bona fide} classical properties), and the system determinately possesses a property if (and only if) it is in the subspace projected onto by the associated projector, and determinately does not possess a property if (and only if) it is orthogonal to that subspace. As such, many properties seem to be neither determinately possessed nor determinately not possessed.

This understanding is adequate for interpretations of quantum mechanics in which the theory is simply an algorithm for predicting measurement outcomes; it is adequate for interpretations which abandon classical logic. For more traditionally realist interpretations---in particular for hidden-variable theories like Bohm's, dynamical-collapse theories like GRW's, or for the Everett interpretation---it is hopeless. Such theories read the quantum state literally, as itself standing \textit{directly} for a part of the ontology of the theory. To every different quantum state corresponds a different concrete way the world is. For Everett and for some readings of dynamical-collapse theories, the quantum state (perhaps together with some background space or spacetime) gives the \emph{only} part of the ontology.

From this perspective, regarding the state as encoding properties of the system in the traditional way is at best unhelpful and incomplete---many properties, like ``being in an entangled state'' or ``being in some eigenstate of energy'' or ``possessing an even number of zero amplitudes in configuration-space'' cannot be expressed using the traditional approach\footnote{E.g., the disjunctive property of being in some energy eigenstate or other would, in projectors language, be given by a sum of projectors onto a complete set of energy eigenstates, returning the projector onto the whole Hilbert space; thus every quantum state has this property. Clearly something has gone wrong.}. Focusing on projectors to represent properties is too crude to capture all of the interesting properties of the world when the quantum state directly describes ways the world is. At its worst, the traditional approach can be actively misleading: for instance, it is responsible for much of the confusion about the ``problem of tails'' in dynamical-collapse theories and makes obscure the nature of the branches in the Everett interpretation. (These claims are defended \emph{in extenso} in \citeN{wallaceQMreview}).

In that case, how \emph{should} we think about the quantum state? Mathematically it is a vector (or, strictly, a ray) rotating in a very high-dimensional Hilbert space. But this feels unsatisfactory: the physical universe is, as we have said, very highly
structured, whereas Hilbert-space vectors seem pretty much alike. The
problem is partially resolved when we recall that a quantum theory can
only be specified if we also give some particular set of `observables' (e.g.,
the position and momentum operators in one-particle QM, or the various field observables in QFT)---this breaks
the symmetry of Hilbert space, allowing us to see how merely mathematically distinct vectors could represent physically distinct states of affairs. (In a somewhat similar way, the complicated structure of a classical-mechanical system is lost when that system is represented by a mere point on phase space, until it is recognised that phase space is a highly structured, non-homogeneous manifold.) The state vector can now be seen as a way
of codifying the various rich properties of the physical state: in one-particle
quantum mechanics, for instance, to any self-adjoint function $f(\op{X},\op{P})$ of
the position and momentum operators it assigns an `expectation value'
$\matel{\psi}{f(\op{X},\op{P})}{\psi}$. Differences between states can now be understood in terms of their differing patterns of assignment of numbers to operators.

But how should such properties be understood? This is where perplexity sets in. As argued above, anyone who takes the quantum
state seriously should avoid
definitionally relating it to measurement: whatever ``expectation
value'' means, it had better not mean the mean result of measuring the
observable. Yet what other sense might we assign to it?


It is at this point that one may turn to the lessons illustrated in Section~\ref{prolegomena} for guidance. Can we follow the example that electromagnetism presented and seek to grasp the meaning of the otherwise alien quantum state by relating it to spatiotemporal properties?

At first blush it might seem not: the problem in quantum mechanics is that the quantum state does not
trivially have a suitable spacetime representation; in particular,
it is not representable as any ordinary sort of spacetime field.  But a natural move is available:
\emph{wave-function realism}, which has been explored recently
by (e.g.) \citeN{albertmetaphysics}, Monton \citeyear{monton2002,montonconfiguration} and
\citeN{lewisconfiguration}. Albert argues, following the lead of \citeN{bellqmforcosmologists}, that the quantum state, if understood physically at
all, should be understood in terms of its configuration-space
representation: that is, as a complex-valued field on 3$N$-dimensional
space, for an $N$-particle quantum theory. In this picture we have clear bearers and properties: the bearers are the points of a space at which certain properties---which happen to be represented by a complex number (or vector, if we include internal degrees of freedom)---are instantiated.
Understood this way, there
seems to be nothing more mysterious about the quantum state than about
any other complex-valued field---except that it lives not on
spacetime, but on a far-larger space---a space, furthermore, which (just as in the example of phase space considered in section~\ref{prolegomena}) is highly non-homogeneous and non-isotropic, since its dynamical symmetry group is far smaller than $SO(3N)$. This seems a perfectly respectible and intelligible candidate for a quantum ontology. (\citeN{vaidmanencyclopedia} and
\citeN{barbour99} describe their respective versions of the Everett
interpretation from the perspective of wave-function realism, though
they do not explore its implications.)

If wave-function realism is correct (and if it alone, and not some hidden variables, is the physical basis for observed reality), the world is really
3$N$-dimensional at its most fundamental level, and our 3-dimensional world
is in some sense emergent from it.

Wave-function realism has not been without its critics, however. Monton~\citeyear{monton2002,montonconfiguration}, for example, raises the questions of whether the wave-function realist ontology could underwrite requisite facts about three-dimensional objects and their behaviour; and whether it provides a setting in which suitable facts about mental goings-on could obtain.

The argument in \cite{monton2002} focuses on whether or not there is a unique mapping from a) points in the 3$N$-dimensional space on which the field represented by the wave-function lives, to b) configurations of $N$ three-dimensional objects. Now, of course we agree that \emph{purely mathematically} any number of such mappings can be constructed, but that is not of prime importance. Rather, in our view, the central question is whether the \textit{structure of the field} that the wave-function represents could (both synchronically and diachronically) provide a supervenience base for facts about the three-dimensional objects and spatial relations among them of our experience; and it seems it can.
Three-dimensional features will emerge as a consequence of the dynamics (in large part due to the process of decoherence): the form of the Hamiltonian will ensure that the wave-function decomposes into components evolving autonomously according to quasi-classical laws. That is, each of the decohering components will correspond to a system of well-localised (in 3-space) wavepackets for macroscopic degrees of freedom which will evolve according to approximately classical laws displaying the familiar three-dimensional symmetries, for all that they are played out on a higher-dimensional space. Three-dimensional quasi-classical structures are thereby recovered (just as in the case of classical mechanics which we mentioned in section~\ref{prolegomena}) and mental events---should one be especially concerned about them---will find their usual home.

Monton might be concerned that this structure of the field could underwrite \textit{more than one} set of emergent 3-spaces, but even if this were so (which seems highly unlikely\footnote{Permuting the 1-dimensional variables in the expression of the wave-function into different groupings of three, as Monton envisages, may well not give rise to structures that are diachronically isomorphic to three-dimensional objects and relations. The ensuing structures certainly wouldn't obey the empirically correct quasi-classical laws.}) it would mean only that more than one set of emergent 3-spaces existed at the same time, supervening on the wave-function ontology, not that none did. (The pattern of thought here is the same as that applied in decoherence-based solutions to the preferred-basis problem in Everett, cf. \citeN{GMH:iguses}; \citeN{saundersdecoherence}; \citeN{wallacestructure}.)

In his 2006, Monton changes tack slightly, arguing that the main problem with wave-function realism is its revisionary nature, rendering false reams of everyday judgements concerning the nature of ordinary objects (e.g., that they are three-dimensional); but again we demur. While the wave-function realist will deny that three-dimensional objects and spatial structures find a place in the fundamental ontology, this is not to say that the three dimensional objects surrounding us, with which we constantly interact, and which we perceive, think and talk about, do not exist; that there are not truths about them; it is just to maintain that they are emergent objects, rather than fundamental ones. But an emergent object is no less real for being emergent. It is also worth keeping in mind that many workers in quantum gravity have long taken seriously the possibility that our four-dimensional spacetime will turn out to be emergent from some underlying reality that is either higher-dimensional (as in the case of string theory\footnote{Popular accounts of string theory can sometimes give the impression that the emergence of four-dimensional spacetime from an underlying 11-dimensional space is purely a matter of coarse-graining, of ignoring sufficiently small distances; the actual physics is significantly more complex, however, and makes essential use of dynamical principles. (For instance, it has been argued that the extra dimensions might have diameters up to several centimetres; and we certainly don't coarse-grain that much!) The correct criterion for whether a given higher-dimensional theory is empirically adequate is, instead, whether it can be approximated at low energies by an effective field theory which can be expressed on a four-dimensional background. We are grateful to an anonymous referee for pressing us on this point.}) or not spatiotemporal at all (as in the case of loop quantum gravity). In neither case is it suggested that ordinary spacetime is \emph{nonexistent}, just that it is \emph{emergent}.

\section{Problems with wave-function realism}\label{problems}

But if wavefunction realism is a coherent metaphysical possibility, we feel it nonetheless sits uneasily with real physics. One of our worries can be understood even in the context of nonrelativistic quantum mechanics (NRQM): namely, physicists normally regard all bases as on a par (so that the configuration-space representation of the quantum state is just one representation among many); they also regard various different formulations of the dynamics (Heisenberg, Schr\"{o}dinger, interaction-picture) as on a par. Both of these are violated by wave-function realism, which gives the Schr\"{o}dinger form of the dynamics, and the position representation, a special status.\footnote{Is this just a sociological observation? Not in the relevant sense. Of course, we do not want to claim that the philosophical commitments of physicists---even those demonstrated in their practice as well as their own philosophising---are \emph{infallible} (the measurement problem is an obvious counter-example). But equally, we feel that philosophers of a naturalistic bent should be cautious in arguing for the metaphysicial priority of any formulation of a physical theory that is unnatural by physics' own standards.}

To be fair, a case can be made in NRQM that both position-basis formulations of the theory, and the Schr\"{o}dinger picture, do indeed have a special status; indeed, many students first learn NRQM in this form, and move on to the Hilbert-space formalism only later. If NRQM were the only quantum theory with which we were concerned, it would \emph{perhaps} not be unreasonable to regard the configuration-space formalism as primary, and the Hilbert-space alternative just as a mathematical curiosity.

This brings us to quantum field theory (QFT), where things are not so straightforward. In our view, two related worries about QFT make wave-function realism an unattractive position at best, an unviable one at worst:
\begin{enumerate}
\item In quantum field theory, the particles are not fundamental and their positions are imprecisely defined; as such, no really satisfactory notion of ``configuration space'' is available for us to formulate wave-function realism.
\item Wave-function realism obscures the role of spacetime in the formulation of quantum theory.
\end{enumerate}
(A third problem, how wave-function realism deals with nonlocality and relativistic covariance, will be postponed until Section~\ref{locality} and following).

To begin with the first problem: in QFT, particle number is not conserved. If there is a ``configuration-space'' representation of the quantum state, it is given by assigning a (non-normalised) wave-function to each of the infinitely many $3N$-dimensional configuration spaces. Wave-function realism now gives us a picture, not of a single field undulating in a high-dimensional space, but of an infinite number of such fields, each undulating in its own high-dimensional space and each interacting with the others.

Is this picture problematic? It is certainly unintuitive; but so was wave-function realism itself, and we have been at pains to stress that this is no criticism of it. But things are actually rather worse than this, in fact. For particles are not only non-conserved in QFT, they are non-fundamental: mathematically speaking they are emergent entities supervenient on an underlying field ontology. In fact, the properties of these entities (mass and charge, for instance) do not take precisely-defined values, but must be chosen according to the particular situation under consideration  (see \citeN{wallaceconceptualqft} for a more detailed discussion). This makes particle configurations unattractive---technically as well as conceptually---as the basis for defining the ontology of QFT.

A superficially attractive alternative is to use a basis of \emph{field} configurations: in vacuum quantum electrodynamics, for instance, the wave-function would assign a complex number to each possible state of the magnetic field. Such a wave-function would be defined on an infinite-dimensional space\footnote{We ignore details of renormalisation; if we regard the high-energy cutoff as real (as advocated by \citeNP{wallaceconceptualqft}) and if the universe is spatially closed then the dimension of the space will be finite but extremely large.} but at least we would only have one such wave-function.

The problem with this move is that there is no single preferred choice of fields by which a QFT can be specified.
A number of results
from QFT (from formal results like Borchers equivalence \cite{haag} to concrete models like the
Sine-Gordon equation \cite[pp.\,246--252]{colemansymmetry}) suggest that a single QFT can be equivalently described in terms
of several different choices of field observable, with nothing in particular to choose between
them.

We might sum up the objection thus: wave-function realism requires a metaphysically preferred basis. In NRQM, there is a single natural choice of such basis: the configuration-space basis is conceptually natural and dynamically special. There is no such natural choice in QFT: many different bases are dynamically on a par and the ``natural'' choice is situation-dependent. This makes any particular metaphysical preference appear somewhat ad hoc.

(This objection is probably most significant for Everettians, who generally regard it as a virtue of their preferred interpretation that it requires no additional formalism, and so are unlikely to look kindly on a requirement in the metaphysics for additional formalism. Advocates of dynamical-collapse and hidden-variable theories are already committed to adding additional formalism, and in fact run into problems in QFT for rather similar reasons: there is no longer a natural choice of basis to use in defining the collapse mechanism or the hidden variables. We are not ourselves sanguine about the prospects of overcoming this problem\footnote{Extant proposals for Bohmian QFTs, in particular, seem to us to pay insufficient attention to the problem of renormalisation, though a full discussion lies beyond the scope of this article.}; but if it \emph{were} to be overcome, the solution might well also suggest a metaphysically preferred basis to use in formulating a QFT version of wave-function realism.)

What of the second objection? We stress that we have no \emph{conceptual} objection to a view of spacetime on which it is emergent from some more fundamental theory; rather, we do not find this a good account of spacetime in physics as we find it. Whilst it is true that spacetime in nonrelativistic quantum mechanics is somewhat elusive, it is mathematically fundamental in QFT: the dynamical variables in QFT are explicitly associated with spacetime regions.

At least heuristically, these dynamical variables can be thought of as being built up from integrals over products of field operators located at a point, such as
\be \op{O}=\int_\Delta \dr{x}\op{\phi}(x)\op{\phi}(x)\ee
where $\Delta$ is some spacetime region. (The point of Borchers equivalence is that \op{O} could be expressed in terms of different field operators whilst still being associated with the same region $\Delta$.)

The objection is sharper still in the case of semiclassical quantum gravity (see, for instance, \citeNP{waldqft}). Here we define the quantum fields on a \emph{curved} spacetime background and require that background to satisfy the classical Einstein field equations with respect to the expected distribution of stress-energy. Such models are used to address issues like black hole evaporation and early-universe cosmology; there are sound theoretical reasons to regard them as an appropriate limiting case to full quantum gravity in some regimes, though there is very little direct empirical evidence for their accuracy. It is opaque to us how semiclassical quantum gravity can be understood except via a quantum ontology which treats spacetime as fundamental.

If wave-function realism were ``the only game in town'', these objections would have to be overcome somehow; by biting various bullets, perhaps they can be. But in fact an alternative ontology is available, one which respects the democracy of Hilbert-space bases and which gives a more central role to spacetime. This alternative, we believe, bears much the same relation to wave-function realism as the $N$-particles-in-space view of classical mechanics bears to the one-particle-in-phase-space view. We present it in the next section.

\section{Spacetime state realism}\label{st ontology}

Wave-function realism presented a clear ontology by preferring a particular set of operators and finding a spatial arena by enlarging the fundamental physical space from three dimensions to 3$N$ dimensions; thereby providing sufficient property bearers. If we do not wish to prefer a basis, then we should stick to the characterisation of the quantum state of a system as a (positive, normalised) linear functional of the dynamical variables; that is, mathematically, as a density operator on the Hilbert space of the system. To go from this to a candidate ontology, though, we still require both clear bearers of properties and a clear view of the structure that a quantum state would ascribe to the world. We propose that what is crucial is an analysis of the total system---the Universe---into subsystems.

Suppose one were to assume that the Universe could naturally be divided into subsystems; assign to each subsystem a density operator\footnote{We intend that if $A$ and $B$ are both subsystems of the Universe then $A\cup B$ is too. On the significance of this, see Section~\ref{locality}.}. We then have a large number of bearers of properties---the subsystems---and the density operator assigned to each represents the intrinsic properties that each subsystem instantiates, just as the field value assigned to each spacetime point in electromagnetism, or the complex number assigned to each point in wave-function realism, represented intrinsic properties. While the property that having a given density operator represents may not be a familiar one, the case need be no different in principle from that of electromagnetism. In so far as one can continue to press for the physical meaning of the density operator, the theory in which these objects are postulated must provide the answer. (We can say such things, for example, as: the property is not a scalar one, in contrast to wave-function realism. We sketch how the relation to the macroscopic goings-on of our experience would be developed in Section~\ref{appearances}.) To provide a simple model, imagine a Universe consisting of a great many interacting qubits (cf. \citeN{deutschhayden} or von Weizsacker's \textit{ur}-alternatives \cite{vonweizsacker}). The qubits each bear the property or properties represented by their two dimensional density operator; pairings of qubits bear properties represented by a four dimensional operator; and so on. There need be no reason to blanche at an ontology merely because the basic properties are represented by such objects: we know of no rule of segregation which states that only those mathematical items to which one is introduced sufficiently early on in the schoolroom get to count as possible representatives of physical quantities, for example!

Note that if, by contrast, we were to treat the Universe just as one big system, with no subsystem decomposition, then we would only have a single property bearer (the Universe as a whole) instantiating a single property (represented by the Universal density operator) and we would lack sufficient articulation to make clear physical meaning of what was presented (as with the one-particle-in-high-dimensional-space view of classical mechanics, one would struggle to see the structure being imputed to the world in this case).

Now, our proposal becomes more concrete when one connects this system-subsystem analysis of the quantum state with
spacetime. To do this we need some notion of the spatial location of a
physical system. In NRQM this is somewhat
complicated: the systems are naturally taken to be particles and
assemblies of particles, and a particle's spatial location is one of its
dynamical properties, not something to be specified \emph{ab initio}.

It is somewhat simpler if we consider the Fock-space formalism of NRQM, where we
allow the number of particles to vary.\footnote{Recall that if \mc{H} is any Hilbert space, the associated symmetric (bosonic) and antisymmetric (fermionic) Fock spaces are
\[\mc{F}_S (\mc{H})=\oplus_{i=0}^\infty \mc{S}(\otimes^i \mc{H})
\]
and
\[\mc{F}_A (\mc{H})=\oplus_{i=0}^\infty \mc{A}(\otimes^i \mc{H})
\]
where $\mc{S}$ and $\mc{A}$ are symmetrisation and antisymmetrisation operators. The distinction between $\mc{F}_S(\mc{H})$ and
$\mc{F}_A(\mc{H})$ plays no part in the argument, and will be ignored in the text.}
On the one-particle Hilbert space of any non-relativistic particle we can define projectors $\op{P}_\Delta$ projecting onto those states with wave-functions having support in spatial region $\Delta$. If $\{\Delta_i\}$ is a partition of real space into measurable subsets and if $\mc{H}_i=\op{P}_{\Delta_i}\mc{H}$ is the Hilbert space built from all states having support in $\Delta_{i}$, then we have $\mc{H} = \oplus_{i}\mc{H}_{i}$ and moreover
\be \mc{F}(\mc{H})=\mc{F}(\oplus_i \mc{H}_i)=\otimes_i \mc{F}(\mc{H}_i),\label{tensor fock}
\ee
so the Hilbert spaces $\mc{F}(\mc{H}_i)$ may be taken as representing the possible states of the subsystem in, or comprising, region $\Delta_i$.\footnote{Intuitive support for the result expressed in eqn.~\ref{tensor fock} can be seen once we recognise that the Fock-space operation is a sort of `exponential' of the Hilbert space (this is clearest from the power-series expansion of the exponential), so that we can write $\mc{F}(\mc{H})=\exp(\mc{H})$. Then equation~\ref{tensor fock} becomes $\exp(\oplus_i \mc{H}_i)=\otimes_i \exp (\mc{H}_i)$. More formally, for each $i$ let $\opad{a}_{i,k}$ be a set of creation operators for states in $\mc{H}_i$. Then it is easy to see (via the observation that creation operators on different Hilbert spaces commute) that states of form
\[\opad{a}_{i_1,k_1} \cdots \opad{a}_{i_n,k_n}\ket{\Omega}\] (where \ket{\Omega} is the vacuum) form a basis both for $\otimes_i \mc{F}(\mc{H}_i)$ and for $\mc{F}(\oplus_i \mc{H}_i)$.}
This means that we can take the regions of space (and their unions) as our basic bearers of properties:
tensor products of states belonging to each region (in general, superpositions of such products) allow us to express the original total state of varying particle number.

Already our presentation is sounding somewhat field-theoretic. To illustrate what is going on, consider a particular region of space $\Delta_{j}$. This region has a Fock space $\mc{F}(\mc{H}_{j})$ whose (pure) basis states can be represented in the form $|n_{1},n_{2},\ldots\rangle$, where $n_{1}$, $n_{2}$ etc. represent the occupation numbers of what we can think of as the available modes of $\Delta_{j}$, that is, the number of excitations in each of some orthogonal set of states of $\mc{H}_{j}$. What we would normally think of in NRQM as a single particle localised in $\Delta_{j}$ will, in this setting, be represented by a singly excited state of $\mc{F}(\mc{H}_{j})$, e.g. $|1,0,0\ldots\rangle$, tensor product with the vacuum state $|0\rangle$ for all the other regions' Fock spaces. In general, then, a single particle (which usually \textit{won't} be localised in some particular region) will be represented by an entangled state composed of a superposition of states each differing from the vacuum only in a small region $\Delta_{i}$:
\[ \ldots|1,0,\ldots\rangle|0\rangle|0\rangle\ldots +  \ldots|0\rangle|1,0\ldots\rangle|0\rangle\ldots + \ldots|0\rangle|0\rangle|1,0,\ldots\rangle\ldots\]
and so on.

Things become simpler still when we move to full quantum field theory. In the algebraic formulation of QFT, we associate to each spacetime region $\mc{R}$ a $C^*$-algebra $\mc{A}(\mc{R})$ of operators, representing the dynamical variables associated to region $\mc{R}$. A state $\rho$ of such a region is a positive linear functional on $\mc{A}(\mc{R})$ (often described in rather instrumentalist terms as giving the expectation value of each element of $\mc{A}(\mc{R})$) and by the Gelfand-Naimark-Segal construction we can associate $\rho$ with a state in a Hilbert space $\mc{H}_R$, and  represent $\mc{A}(\mc{R})$ as an algebra of operators on $\mc{H}_R$ (see, \egc, \citeN[pp.122-124]{haag} for the details). $\mc{H}_R$ can then be taken as the Hilbert space of the field in region $\mc{R}$.\footnote{In the standard presentations of AQFT, the algebra $\mc{A}(\mc{R})$ is infinite-dimensional; as such it has multiple non-isomorphic representations, and so different states lead to different Hilbert spaces. This makes it less clear that we are licensed to talk about $\mc{H}_R$ as ``the'' Hilbert space for region $\mc{R}$. One of us has argued elsewhere, however \cite{wallaceconceptualqft} that this is an artefact of the formalism, which disappears when we properly understand the renormalisation process; as such, we ignore this complication in the text.} If preferred, one can even remain at the more abstract level, forego the representation theorems and just take the $C^{*}$-algebraic state itself as denoting the properties of a region.

This alternative ontology avoids the problems previously identified for wave-function realism: it is well-defined for any quantum theory with compositional structure (in particular, for any quantum field theory, and for any many-particle theory once it is expressed in field-theoretic terms); it respects the dynamical structure of QM, indicating no preference for Schr\"{o}dinger over Heisenberg or interaction dynamics (as the state is just construed as a linear functional of the dynamical variables); it adds no additional interpretational structure (given that the compositional structure of the system is, \emph{ex hypothesi}, already contained within the formalism); and it gives an appropriately central role to spacetime. For want of a better name, we call it \emph{spacetime state realism}.\footnote{Although if the theory has a compositional structure not induced by spacetime regions---as in, say, an abstractly specified quantum computer---this name is a misnomer.}

\section{Locality}\label{locality}

How is one to visualise the quantum state, according to our proposal? Recall the discussion of Section~\ref{prolegomena}: typically in classical physics we understand physical systems via the properties they ascribe to spacetime regions. Wave-function realism keeps this basic framework at the cost of greatly enlarging the background spacetime: the quantum state is a complex field much like the classical Klein-Gordon field, but it is defined on an extremely high-dimensional space. Our proposal is in one sense closer to the classical situation: we associate a set of properties (represented by a density operator) to each region of spacetime. In another sense, though, it is less classical than wave-function realism: it displays a certain sort of nonlocality.

To see this, let $A$ and $B$ be spacelike separated regions. The universal quantum state, like a classical field, associates a state to $A$, $B$, and $A \cup B$. But there is an important disanalogy. Classically, the separate states of $A$ and $B$ completely determine the state of $A \cup B$---once the electromagnetic field, say, is known in $A$ and $B$, it is known also in $A \cup B$. \citeN{healeyseparability} calls this property \emph{separability}: the principle that the properties of a spacetime region supervene on the properties of its spatiotemporal parts (see also \cite{earman:locality} for a discussion of various senses of locality). We might equally call it \emph{Humean supervenience}, after Lewis's  thesis---which he attributes to Hume---that all properties of the world supervene on properties of spacetime points and on the relations between those points \cite[pp.ix--xvii]{lewispapers2}.

Thanks to the phenomenon of entanglement, separability is simply false for the quantum state according to spacetime state realism: if the state $\denop_{A\cup B}$ is known, then via the partial trace operation we can learn the states of $A$ and $B$, but of course the converse is not so.

We should not be surprised, perhaps: quantum mechanics is nonlocal; everyone knows that, don't they? But notice that the description given of nonlocality is very different according to wave-function realism. There, the state \emph{on configuration space} is completely separable: specify a complex number at each point and you specify the state. According to wave-function realism, non-separability on three-dimensional space is an illusion, or better, an emergent property---it is a consequence of our attempt to use three-dimensional language to describe an entity which fundamentally lives in $3N$-dimensional space. By contrast, in spacetime state realism non-separability is fundamental.

Is this a reason in favour of wave-function realism (setting aside our previous arguments against its tenabity)? Not in our opinion, for three reasons:

\begin{enumerate}
\item As nonlocal forms of behaviour go, non-separability is fairly mild. It does not imply any sort of action at a distance: the quantum state of spacetime region $A$ is dynamically determined by the state of its past light cone (more precisely: by the state of any spacelike slice of its past light cone). The state of $A\cup B$ may indeed be changed by operations in the vicinity of either $A$ or $B$, but the state of $B$ is unaffected by operations performed at $A$.
(Note also that it is milder than the nonlocality predicted by the Bell inequalities: those concern changes made to the local state at $A$ as a consequence of operations performed at $B$. Since we are considering only unitary quantum mechanics---and so, by implication, presuming the Everett interpretation---some of the premises of Bell's theorem aren't in play. (See e.g. \citeN{timpsonbrown:er} for a discussion of Bell's theorem and locality in Everett.))

We are, in fact, unmoved by purely \emph{metaphysical} objections to action at a distance. Our concern about it stems from its tension with relativistic covariance. But there is no such tension in the case of non-separability.

\item It is tempting to regard separability as part of our ordinary conception of space: arguably, if some putative spacetime has essentially nonlocal properties, or perhaps better, if the things in it (e.g., fields) have to end up being attributed non-separable properties, we ought not to call the arena ``spacetime''.\footnote{Essentially this idea may be found in Einstein: ``It is...characteristic of...physical objects that they are thought of as arranged in a  space-time continuum. An essential aspect of this arrangement of things in physics is that they lay claim, at a certain time, to an existence independent of one another, provided these objects `are situated in different parts of space'. Unless one makes this kind of assumption about the independence of the existence (the `being thus') of objects which are far apart from one another in space...physical thinking in the familiar sense would not be possible." \cite{einstein1948}} But there is nothing \emph{mathematically} improper about these nonlocal properties. And even according to wave-function realism, the four-dimensional spacetime we live in (an emergent object, supervenient on the wave-function) itself involves non-separability in exactly the way described by spacetime state realism. The two ontologies are not in disagreement about whether ordinary spacetime allows \emph{separability}; they are in disagreement about whether this spacetime is \emph{fundamental}. And if our paradigm example of a space involves non-separability, we see no plausible argument that our conception of space \emph{requires} separability. Rather, we have a mere intuition that it does---an intuition developed because space \textit{looks} separable, unless you look very carefully.

\item In any case, there are respectable \emph{classical} examples of non-separability. In particular, the Aharonov-Bohm effect is best understood as implying that electromagnetism is non-separable: specifying the connection on regions $A$ and $B$ does not suffice to determine it on $A\cup B$ unless the latter region is simply connected. If non-separability is reasonable for the A-B effect, why not for the quantum state?\footnote{Why do we call the A-B effect classical? Because mathematically it is a classical consequence of the interactions between a $U(1)$ connection and a complex field. Inconveniently, the complex fields we see around us (the proton and electron fields) have no accessible classical limit, so we observe its effect on the quanta of those fields, but the effect is not \emph{essentially} quantum-mechanical. See \citeN{wallaceantimatter} for more on this point.}

(Of course, the implications of the A-B effect are not uncontroversial; one might as easily argue that non-separability is unreasonable in both cases. We expand on our preferred reading of the A-B effect, and on the analogies between that effect and quantum non-locality, in \citeN{part2}).

\end{enumerate}

In fact, once we consider the relativistic domain there are active \emph{advantages} to spacetime state realism's treatment of nonlocality, as we will see in Section~\ref{relativity}. First, however, it will be useful to consider how the spacetime state ontology fares in the case of non-unitary dynamics.

\section{Non-unitary dynamics on the spacetime ontology}\label{nonunitary}

So far we have been considering quantum states evolving under purely \emph{unitary} dynamics. There are, however, some reasons to be interested in adding non-unitary processes to the dynamics---specifically, if we are unhappy with the Everett interpretation and with hidden-variable theories, we may wish to regard wave-function collapse as a real microphysical process and to model it non-unitarily.

It is reasonable to assume that wave-function collapse must be implemented by a non-Lorentz-covariant dynamics, for familiar reasons: if one prepares two particles in an entangled state, separates them widely, and measures one of them, the other is supposed to collapse \emph{immediately}---and ``immediately'' seems to imply a preferred frame. But there is an alternative tradition (defended at various times by, \egc, \citeN{aharonovalbertHDQM}, \citeN{flemingHDQM}, Myrvold \citeyear{myrvoldcollapse,myrvold:becoming}, and \citeN{tumulka}): \emph{foliation-relative collapse} (FRC). According to FRC, a collapse event should be modelled as occurring not \emph{at a point} but \emph{on a hypersurface}.

It is difficult to make a really thorough evaluation of the merits of FRC, since it remains a framework only: there is at present no satisfactorily worked-through relativistic collapse theory, let alone one which reproduces the predictions of QFT. Yet two things seem clear. Firstly, FRC does seem to offer, in principle, a framework for a covariant quantum theory which nonetheless incorporates dynamical collapse. Secondly, it has some extremely counter-intuitive consequences.

\begin{figure}
\begin{center}
\rotatebox{-90}{\scalebox{0.8}{\includegraphics{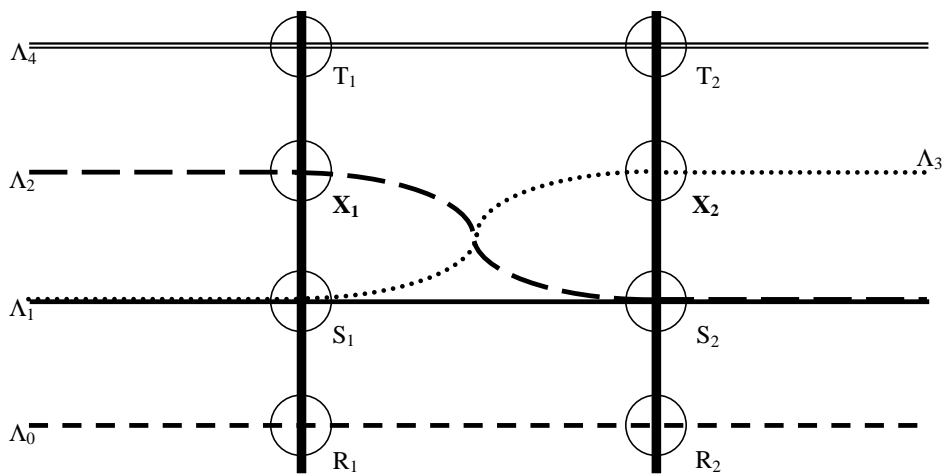}}}
\caption{\small A spacetime diagram of the histories of two particles; time runs vertically. $\Lambda_{0}$ to $\Lambda_{4}$ label distinct hypersurfaces of simultaneity. Particles 1 and 2 are spacelike separated and begin (on $\Lambda_{0}$) in an entangled state. Regions $R_{j},S_{j}, T_{j}$ lie on the worldline of particle $j$ (1 or 2). At $X_{1}$, particle 1 is measured; at $X_{2}$, particle 2 is measured. The story one tells about the effects of these measurements is very different depending on whether one uses a foliation in which $X_{1}$ precedes $X_{2}$ ($\{\Lambda_{0}, \Lambda_{1}, \Lambda_{2}, \Lambda_{4}\}$) or in which $X_{2}$ precedes $X_{1}$ ($\{\Lambda_{0}, \Lambda_{1}, \Lambda_{3}, \Lambda_{4}\}$). Note that in the spacetime state framework, we may replace talk of states of particles relative to elements of a foliation with direct talk of the states of spacetime regions, such as $R_{j},S_{j}$ and $T_{j}$.}\label{foliations}
\end{center}
\end{figure}

To illustrate this latter point, suppose that we prepare two widely-separated particles in an entangled state (see Figure~\ref{foliations}), so that on surface $\Lambda_1$ the combined state is
\be \ket{\psi_1}=\alpha\ket{0}\ket{1}+\beta\ket{1}\ket{0};\ee
and we go on to measure the spins of both of them (with the measurement on particle $i$ occurring at point $X_i$).

If we analyse the situation using a foliation containing $\Lambda_1$, $\Lambda_2$ and $\Lambda_4$---that is, on a foliation in which particle 1 is measured first and particle 2 second---the history of the system will be something like: first there is a stochastic process (between $\Lambda_1$ and $\Lambda_2$) caused by the measurement of particle 1, in which particle 1 collapses into (say) $\ket{0}$ and particle 2 into \ket{1}; then there is a deterministic process (between $\Lambda_2$ and $\Lambda_4$) where we passively determine that the state of particle 2 is indeed in state \ket{1}. However, the history looks very different on a foliation containing $\Lambda_1$, $\Lambda_3$ and $\Lambda_4$, so that 2 is measured before 1. According to this second foliation, the stochastic process (between $\Lambda_1$ and $\Lambda_3$) is triggered by the measurement of particle 2, and it is the measurement of particle 1 (between $\Lambda_3$ and $\Lambda_4$) that is the passive observation.

Note, however, that the \emph{observed consequences} of the two measurements are the same (particle 1 has spin up; particle 2 has spin down), and that the end state on $\Lambda_4$ is the same in both foliations,
\be \ket{\psi_4}=\ket{0}\ket{1}.\ee
This makes it a somewhat subtle business to say just what, if anything, is unacceptable about FRC. Indeed, some have argued that in fact there is nothing wrong with it, and that its consequences---though indeed counter-intuitive---are inevitable given relativity and non-locality, and should not be regarded as unacceptable. (See Myrvold \citeyear{myrvoldcollapse,myrvold:becoming} and references therein.)

We are unpersuaded, but we do not want to defend this here. Our goal in this paper is more modest: to use the spacetime state framework to clarify the sense in which non-unitary dynamics has consequences going beyond those of unitary dynamics. To this end we shall now allow ourselves to talk directly of the states of the pertinent spacetime regions shown in Figure~\ref{foliations}; considering first the unitary case, before seeing how things differ under non-unitary dynamics.

Thus suppose that in Figure~\ref{foliations} the transformations performed at $X_1$ and $X_2$ were unitary---say,
\[\ket{0}\longrightarrow \ket{1}\]
\be
\ket{1} \longrightarrow -\ket{0};
\ee
a spin-flip rather than a measurement. The states of the various relevant regions are easily calculated:
\[
\denop_{R_1}=\denop_{S_1}=|\alpha|^2 \proj{0}{0}+|\beta|^2 \proj{1}{1}\]\[
\denop_{R_2}=\denop_{S_2}=|\beta|^2 \proj{0}{0}+|\alpha|^2 \proj{1}{1}\]\[
\denop_{T_1}=|\beta|^2 \proj{0}{0}+|\alpha|^2 \proj{1}{1}\]\[
\denop_{T_2}=|\alpha|^2 \proj{0}{0}+|\beta|^2 \proj{1}{1}
\]
and
$\denop_{\Lambda_i}=\proj{\psi_i}{\psi_i}$
where
\[\ket{\psi_1}=\alpha\ket{0}\ket{1}+\beta\ket{1}\ket{0}
\]
\[\ket{\psi_2}=\alpha\ket{1}\ket{1}-\beta\ket{0}\ket{0}
\]
\[\ket{\psi_3}=-\alpha\ket{0}\ket{0}+\beta\ket{1}\ket{1}
\]
\[\ket{\psi_4}=-\alpha\ket{1}\ket{0}-\beta\ket{0}\ket{1}.
\]
We observe that, as expected, the state on both $\Lambda_2$ and $\Lambda_3$ is changed from $\Lambda_1$, since either  $X_1$ or $X_2$ lies in the past light cone of each region. The states of the individual particles, however, are unambiguously fixed by the dynamics, and neither particle's \emph{individual} state is affected by a transformation outside its light cone --- $X_1$ lies outside the past light cone of $S_2$, for instance, so $\denop_{S_2}=\denop_{R_2}$ even though $S_2$ is part of $\Lambda_2$ and the state of $\Lambda_2$ is changed by $X_1$.

Things are otherwise in the event of collapse, where we make a measurement rather than a spin flip at $X_i$. The (hypothetical) stochastic dynamics produce perfectly reasonable (non-deterministic) predictions for all the joint states---assigning a non-zero probability to, say, $\denop_{\Lambda_i}=\proj{\phi_i}{\phi_i}$ where
\[\ket{\phi_1}=\alpha\ket{0}\ket{1}+\beta\ket{1}\ket{0}
\]
\[\ket{\phi_2}=\ket{\phi_3}=\ket{\phi_4}=\ket{0}\ket{1}.\]
But there is no unique prescription for the one-particle states. Take $S_2$, for instance---if it is regarded as part of $\Lambda_1$ then it has state
\be |\beta|^2 \proj{0}{0}+|\alpha|^2\proj{1}{1};\ee
if it is regarded as part of $\Lambda_2$ then it has state $\proj{1}{1}$.  We might interpret this (at the risk of confusion with a terminology adopted elsewhere in philosophy of QM) as a \emph{contextuality} of local states---the state of a region depends on which larger region it is regarded as a part of. We might also interpret it as denying the existence of states for many regions of spacetime---regions only have states at all if they are assigned the same state by all foliations. We might call this \emph{nihilism} about local states, a name which will be particularly apt if, as we are inclined to suspect (but have not proved), in realistic models virtually no regions smaller than entire foliations are assigned states.

To conclude this section: there is an ``hierarchy of counterintuitiveness'' here. We have
\begin{description}
\item[1. Full locality] (classical physics). All spacetime regions have states; the state of a region supervenes on the states of its subregions.
\item[2. Non-separability] (unitary quantum physics). All spacetime regions have states but the state of a region does not supervene on the states of its subregions.
\item[3a. Contextuality] (non-unitary quantum mechanics, first interpretation). Spacetime regions smaller than entire hypersurfaces have states which are dependent on which larger region they are regarded as part of.
\item[3b. Nihilism] (non-unitary quantum mechanics, second interpretation). Some or all spacetime regions smaller than entire hypersurfaces have no states at all.
\end{description}

\section{Relativity and Poincar\'{e} covariance}\label{relativity}

Relativistic covariance is straightforward to understand according to spacetime state realism: if $A$ is a spacetime region and $\mc{L}$ is a Poincar\'{e} transformation (represented on Hilbert space by some operator $\op{U}_\mc{L}$) then the state of region $A$ after performing $\mc{L}$ as a global active transformation is determined by the state of $\mc{L}^{-1} A$ before the transformation. This is easy to see: the state of $A$ is just given by the expectation value of all operators localised in $A$. So after the active transformation, to specify the state we need terms like
\be \matel{\psi}{\opad{U}_\mc{L} \op{\Phi}(x)\op{\Phi}(y) \op{U}_\mc{L}}{\psi}\ee
where $x$ and $y$ are points in $A$. But if $\op{\Phi}$ is a scalar field, by the construction of a QFT we have
\be   \opad{U}_\mc{L} \op{\Phi}(x)\op{U}_\mc{L}=\op{\Phi}(\mc{L}^{-1}x),
\ee
and so
\be \matel{\psi}{\opad{U}_\mc{L} \op{\Phi}(x)\op{\Phi}(y) \op{U}_\mc{L}}{\psi}
= \matel{\psi}{\op{\Phi}(\mc{L}^{-1}x)\op{\Phi}(\mc{L}^{-1}y) }{\psi};\ee
considering vector or spinor fields adds only technical complexity.

The situation is rather different for wave-function realism. There is a sense in which the theory is Poincar\'{e}-covariant---namely, that there is a representation of the Poincar\'{e} group as unitary operators on the space of wave-functions on the configuration space\footnote{Those readers who recall our criticisms of wave-function realism in Section~\ref{problems} may reasonably ask: which configuration space? Two natural answers are available: if we consider a fixed number of particles which do not interact, we can use `ordinary' $N$-particle quantum mechanics; more generally we can use a field configuration space.} such that a sequence of wave-functions $\Psi(t)$ satisfy the Schr\"{o}dinger equation only if the sequence $\op{U}_\mc{L}\Psi(t)$ does for all $\mc{L}$. (Put another way: the Schr\"{o}dinger equation is form-invariant under the action of this representation of the Poincar\'{e} group.) But since the wave-function is a function on a space of dimension far higher than three, it is somewhat opaque what the connections are between this result and the usual interpretation of the Poincar\'{e} group in terms of the symmetries of Minkowski spacetime\footnote{A similar concern is raised by \citeN{lewisconfiguration}, who queries the apparent mismatch between the symmetry of the Hamiltonian and the structure of the space on which the wavefunction is defined in NRQM.}.

The problem may be sharpened via a remarkable observation due to David Albert \cite{albertnarrativity}. If wave-function realism is correct, then we might naturally expect that the entire history of the universe should be given by the sequence $\Psi(t)$ for all $t$. As in the classical case, an alternative sequence like $\op{U}_\mc{L}\Psi(t)$ should be a mere redescription of the first sequence. But in fact, this is not the case, as can be shown by a simple example. Consider again the situation of Figure~\ref{foliations}: two distinguishable spin-half particles $1$ and $2$ are prepared in a singlet state\footnote{This is a simplified version of Albert's own example, which involves four particles. Albert's example, however, has the advantage of being manifestly covariant, demonstrating that the non-covariance of Scenario B is not important.}
\be
\ket{\psi}=\frac{1}{\sqrt{2}}(\ket{0}\ket{1}-\ket{1}\ket{0}).
\ee
We now consider:
\begin{description}
\item[Scenario A:] The spin degrees of freedom of the particles remain unaltered for all time.
\item[Scenario B:] At $X_1$ and $X_2$ we perform a spin flip operation on $1$ and $2$ respectively:
\[\ket{0}\longrightarrow \ket{1}\]
\be
\ket{1} \longrightarrow -\ket{0}.
\ee
\end{description}
If we describe the evolution of the state in a foliation where $X_1$ and $X_2$ are simultaneous (one containing $\Lambda_0$, $\Lambda_1$, $\Lambda_4$, say), then scenarios A and B have the same effect on the quantum state: nothing. That is, the state remains a singlet at all times.

This is not the case in other foliations. Consider a foliation where $X_1$ precedes $X_2$ (i.e., we include $\Lambda_{2}$ of Figure~\ref{foliations}). In scenario A, of course, the state remains unchanged at all times. But in scenario B, there is a period during which 1 but not 2 has undergone the spin flip, leading to a global state
\be\ket{\phi^{+}}=\frac{1}{\sqrt{2}}(\ket{1}\ket{1}+\ket{0}\ket{0}).\ee

It follows from this that the sequence $\ket{\Psi(t)}$ does not actually fix the Lorentz-transformed sequence: two worlds may be described by the same sequence but have different Hamiltonians, and so will have different descriptions following a Lorentz transformation. Two observations immediately follow:
\begin{enumerate}
\item For a Poincar\'{e}-covariant quantum theory, the sequence of states $\op{U}_\mc{L}\ket{\Psi(t)}$ cannot be interpreted as a mere redescription of the sequence $\ket{\Psi(t)}$.
\item For a Poincar\'{e}-covariant quantum theory, the sequence of states $\ket{\Psi(t)}$ cannot be regarded as completely specifying the properties of the system.
\end{enumerate}
Albert calls a theory \emph{narratable} if specifying a system's state at all times is sufficient to specify all properties of a system. Poincar\'{e}-covariant quantum mechanics is not narratable: if we give the state at all times on a given foliation, we have given something less than the complete description of the system.\footnote{The phenomenon of narratability failure in Poincar\'{e}-covariant quantum theories relates to the fact that, unlike in the Galilean case, the operators generating the Lorentz-boosts depend on the Hamiltonian of the system; see for example \cite{fleming:note}. We thank Jeremy Butterfield for the observation and Wayne Myrvold for the reference.}

One of the common arguments used in favour of the Everett interpretation over other interpretations is that it is fully compatible with special relativity. The conclusion Albert draws from narratability failure is that this presumed advantage is overstated. On the one hand, as we have seen in Section~\ref{nonunitary}, there are Lorentz-covariant frameworks for dynamical-collapse theories---they have counter-intuitive consequences, but Albert interprets these as special cases of the failure of narratability. If he is correct, then if narratability failure is also acceptable in the Everett interpretation, why not in dynamical-collapse theories?

On the other hand, if narratability failure is \emph{not} acceptable, the only alternative is to give up on Lorentz covariance as fundamental and accept a preferred (albeit undetectable) foliation. But if \emph{this} is acceptable in the Everett interpretation, why not in other interpretations (notably, hidden-variable theories like the Bohm theory, or non-covariant collapse theories)?

We reject the first horn of Albert's dilemma. As we have tried to show in Section~\ref{nonunitary}, the consequences of covariant non-unitary dynamics are relevantly worse than those of covariant unitary dynamics. But we wish to draw a different conclusion from Albert's argument: namely, that it provides a further argument for spacetime state realism over wave-function realism.

To elaborate: according to wave-function realism the world is fundamentally local, but the dimensionality of space is very large ($3N$-dimensional)---the state of the universe at an instant is represented by a complex function on that space. Then it is hard to see how to think about the entire history of the universe other than as a complex function on a $3N+1$-dimensional spacetime. And \emph{that} is what narratability failure rules out.

On the other hand, spacetime state realism is fundamentally \emph{non}-local (non-separable, to be more exact).
However, suppose that we have \emph{any} spacetime theory which is (a) non-separable, so that there can be simultaneous spacetime regions $A$ and $B$ such that the state of $A \cup B$ is not determined by the states of $A$ and $B$ separately, and (b) is also covariant. Covariance entails that there can also be \emph{non}-simultaneous spacetime regions whose joint state is not fixed by their separate states. This opens up the possibility of failure of narratability: specification of global states on elements of one foliation on their own will not in general fix the joint states of non-simultaneous regions (regions which do not lie on some single hypersurface in the foliation), hence will not fix the global states on another foliation (as the latter foliation  will involve linking up hypersurfaces which are non-simultaneous, according to the first).\footnote{It is an interesting question exactly to what extent narratability failure is generic in non-separable relativistic theories, quantum or otherwise. Note that if one were able to \text{infer} the dynamics by perusing the sequence of states on a given foliation then this would allow one to calculate the state on any hypersurface from a previous one. So examples illustrating narratability failure will need to exploit the presence of symmetries precluding such inference. We thank Wayne Myrvold and Tim Maudlin for discussion of this point.}

From this perspective, we can explain any  initial surprise at narratability failure as due to our failure to take non-separability sufficiently seriously. In the Good Old Days of classical physics, we thought that to specify the state of a spacetime region it sufficed to specify the state of all arbitrarily small subregions of that region. When confronted with nonseparability, we accepted that this was insufficient, but assumed without justification that it would suffice to specify the state of all elements in an arbitrary \emph{foliation} of the region. In fact, in a nonseparable theory there are no shortcuts. In principle at least, the state of any region must be specified \emph{directly}: no attempt to specify it in terms of any non-trivial decomposition into subregions will work.

This is not to downplay the importance of foliations, of course. For a variety of reasons\footnote{Perhaps best understood in terms of the structure of the decoherence-defined branching in the Everett interpretation, which is in turn determined by the local nature of the dynamics.} we are often only interested in local properties---properties of spacetime points, in the limit. And all of \emph{these} properties are indeed fully specified by the quantum state on each element of a foliation. Furthermore, they are dynamically determined---indeed, all properties, however nonlocal, are dynamically determined---by the quantum state on a \emph{single} element of this or any other foliation, once the Hamiltonian is taken as fixed. But this dynamical observation should not be confused with the entirely different---and quite false---claim that Poincar\'{e}-covariant QM is narratable.

\section{Finding the appearances}\label{appearances}

A final point should be addressed briefly before concluding.
We noted amongst the prolegomena that when presenting an ontology, one hopes to be able to recount, even if only sketchily, how the property bearers and the properties one postulates in the ontology relate to appearances. How might such a story go in the case of spacetime state realism?

The \textit{general} story will be the same as that in any approach which takes unitary quantum mechanics to be a complete theory: decoherence will provide an emergent classicality; more precisely, a superposition of effectively autonomous quasi-classical sequences of events. Observers able to monitor and interact with their surroundings emerge as inhabitants within such independent quasi-classical histories; corresponding to particular complex arrangements of fundamental particles; evolved to take advantage of the existence of enduring records that decoherence allows \cite{GMH:iguses,saundersevolution}.

The only addition to this general picture in our proposal is that the states of spacetime regions are the primary elements.
Determinate quasi-classical goings on will correspond to certain \emph{special} states of large regions (which will themselves correspond (in general) to highly entangled states of the component sub-regions). In the absence of dynamical collapse, though, the states of large regions will typically correspond to no (single) quasi-classical situation. Rather, decoherence will guarantee that the states of these regions are convex \emph{sums} of the special ``quasi-classical'' states:
\be
\denop_A = \sum_i p_i \denop^i_A
\ee
where each $\denop^i_A$ is a quasi-classical state of $A$. Each quasi-classical component evolves independently of the others, so $\denop_A$ encodes not one, but a great many simultaneously present quasi-classical situations. And the spatial extension of the quasi-classical goings-on is encoded in the entanglement between states: if $B$ is some largish region adjacent to $A$, then the state of $A \cup B$ might be
\be\label{region correlation}
\denop_{A \cup B}=\sum_i p_{i} \denop^i_A \otimes \denop^i_B;
\ee
the local states of $A$ and $B$ each encode many simultaneously present quasi-classical situations, and the nonlocal information in $\denop_{A \cup B}$ encodes the connections between them that justifies the ``many-worlds'' language customarily used to describe the Everett interpretation\footnote{The state given by (\ref{region correlation}) is actually what would be termed a `classically correlated' or separable state of $A \cup B$, but this shouldn't mislead one into thinking that the regions are unentangled. Whether or not reduced states of subsytems of a $N$-party entangled state are themselves entangled depends on the type of $N$-party entanglement that obtains. If the global state is pure, then a state like (\ref{region correlation}) for subregions can only arise because of the existence of entanglement.}.
This brings home the point that the true state of a spatial region is very far from being directly accessible to any realistic agent. An observer in region $A$ (present in the quasi-classical situation encoded by $\denop^i_A$, say) might very well speak of \textit{the} state of $A$ being $\denop^i_A$ and \textit{the} state of $B$ as being $\denop^i_B$, but these would be emergent and approximate notions (somewhat akin to Everett's original `relative states'). The true, ontologically primary, state of $A$ would still be $\denop_A$.

To help relate this rather abstract account to the more familiar, if somewhat specialised, examples one is used-to from quantum mechanical measurement theory, we can consider how an idealised measurement situation might look. In general, such cases of measurement would involve the establishment of entanglement between the regions where the object system may be thought of as located and the regions corresponding to the possible positions of macroscopic pointer variables. (The positions of pointers, and later, the states of observers, would constitute parts of the quasi-classical situations described by certain of the quasi-classical states, in the terminology of the preceding paragraph.)  To take a very simple example in illustration, consider the Coleman-Hepp model of a spin measurement \cite{hepp}, discussed by Bell \cite{belloncoleman-hepp}.

The model consists of a single moving spin-1/2 system which interacts with a one-dimensional (semi-infinite) array of spatially fixed qubits as it moves along the array (one can think of this as a one-dimensional version of a Stern-Gerlach measurement). The qubits fixed at their locations $x_{1},x_{2},\ldots,x_{n},\ldots$ all begin with spin-up in the $z$-direction; when the moving spin passes by one of the fixed qubits, the two interact by a controlled-not operation; that is, the qubit's value is flipped if the moving spin is spin-down and is unchanged if it is spin-up (the flip will be completed by the time the moving spin moves on to the next position in the chain). At $t_{0}$, the moving spin starts at $x_{1}$; we can represent the states of each of the regions $x_{i}$ by number states $|{n_{\uparrow},n_{\downarrow}}\rangle_{i}$ for spin in the $z$-direction. If the moving spin begins in an equal superposition of spin-up and spin-down, then the state of the array at $t_{0}$ will be:
\[ \frac{1}{\sqrt{2}}\bigl(|2,0\rangle_{1}+|1,1\rangle_{1}\bigr)|1,0\rangle_{2}|1,0\rangle_{3}\ldots|1,0\rangle_{n}\ldots,\]
that is, a product state of all the regions in the array. At $t_{1}$, however, as the moving spin has passed on to position $x_{2}$, the state becomes:
\[\frac{1}{\sqrt{2}}\Bigl(|1,0\rangle_{1}|2,0\rangle_{2} + |0,1\rangle_{1}|1,1\rangle_{2}\Bigr)|1,0\rangle_{3}\ldots|1,0\rangle_{n}\ldots;\]
regions 1 and 2 are now entangled; and so it continues, until at $t_{n}$ all the regions up to and including the $n$th are entangled together:
\[\frac{1}{\sqrt{2}}\Bigl(|1,0\rangle_{1}|1,0\rangle_{2}|1,0\rangle_{3}\ldots|2,0\rangle_{n} + |0,1\rangle_{1}|0,1\rangle_{2}|0,1\rangle_{3}\ldots|1,1\rangle_{n}\Bigr)|1,0\rangle_{n+1}\ldots.\]
For sufficiently large $n$, the magnetisation of the array (the average of the dipole moments of the spins) will count as a macroscopic variable recording the value of spin possessed by the measured system. If we were to take the union of the regions $x_{1}$ to $x_{n-1}$ at $t_{n}$ as our region $A$, for example, and $\op{P}_{1}$ and $\op{P}_{-1}$ to be projectors onto states with magnetisation of $A$ close to 1 and -1 respectively, then the state of $A$ will be:
\begin{equation}
\denop_{A}=\frac{1}{2}\op{P}_{1} + \frac{1}{2}\op{P}_{-1};
\end{equation}
corresponding to separate, determinate, values of the pointer variable.

\section{Conclusion}

We have seen that a superior alternative to wave-function realism is available when trying to understand the nature of the quantum state. While both the approaches we have discussed exploit the gains in intelligibility that finding a spatial arena confers, by choosing \textit{spacetime} as the arena, we gain a univocal ontological picture for all standard types of quantum theory (non-relativistic fixed particle number; non-relativistic variable particle number; relativistic; field-theoretic) and we avoid artificially picking out certain classes of observables as preferred. While the kinds of property that will be attributed to a spacetime region by the assignment to it of a density operator are not ones with which we are terribly familiar, we have suggested that this kind of feature is generic in fundamental physics: the case, we have suggested, does not differ in principle from the unfamiliarity of the electromagnetic field. Our understanding of the nature of these properties will, in the end, have to be mediated by our becoming familiar with their role within the theory. We sketched, finally, an---admittedly highly schematic---account of the kind of relation these properties will have to experience; a form of account which is familiar from recent discussions of Everettian quantum mechanics.

On the matter of separability, we have suggested that the gains for the wave-function realist position of being separable at the fundamental level are not substantial. Non-separable theories seem perfectly intelligible; and when it comes to considering relativistic covariance, the fact that wave-function realism fails narratabilty looks mysterious and problematic. By contrast, with a non-separable spacetime picture, narratability failure is rendered innocuous as a perfectly natural consequence of adopting a covariant non-separable theory.

On a final note, recall the hierarchy of counterintuitiveness presented in Section~\ref{nonunitary}. Sticking with purely unitary quantum mechanics keeps us in the top-half of the heirarchy. But given all of our experience with the notion of entanglement, we surely can't expect to move all the way to the top and recover the full locality of classical physics, while sticking with a theory empirically equivalent to quantum mechanics. Or can we? This is the question we take up in Part II.

\begin{flushright}
Balliol College and Faculty of Philosophy,\\ University of Oxford,\\ OX1 3BJ, UK.\\ 
david.wallace@balliol.ox.ac.uk\\

Brasenose College and Faculty of Philosophy\\ University of Oxford,\\ OX1 4AJ, UK.\\
christopher.timpson@bnc.ox.ac.uk
\end{flushright}

\section*{Funding}

Arts and Humanities Research Council, CGT.

\section*{Acknowledgements}

We would like to thank Jeremy Butterfield, David Deutsch, Tim Maudlin, Wayne Myrvold and the anonymous referees for helpful comments and discussion. CGT would also in particular like to thank Joseph Melia for helpful discussion.



\end{document}